\documentclass[proof]{pasj00}
\Received{2014 September 30}
\Accepted{2014 November 25}
\SetRunningHead{Recombining Plasma in G290.1$-$0.8}{Kamitsukasa et al.}

\usepackage{times}
\usepackage{lscape}
\usepackage[dvips]{graphicx}
\renewcommand{\arraystretch}{1}

\begin{document}

\title{Global Distribution of Ionizing and Recombining Plasmas in the Supernova Remnant G290.1$-$0.8}
\author{Fumiyoshi \textsc{Kamitsukasa},\altaffilmark{1}
        Katsuji \textsc{Koyama},\altaffilmark{1,2}
        Hiroyuki \textsc{Uchida} \altaffilmark{2}
        Hiroshi \textsc{Nakajima},\altaffilmark{1}
        Kiyoshi \textsc{Hayashida},\altaffilmark{1}
        Koji \textsc{Mori},\altaffilmark{3}
        Satoru \textsc{Katsuda},\altaffilmark{4}
        and Hiroshi \textsc{Tsunemi},\altaffilmark{1}}
\altaffiltext{1}{%
   Department of Earth and Space Science, Osaka University, 1-1 Machikaneyama-cho,
   Toyonaka, Osaka 560-0043, Japan}
\altaffiltext{2}{%
   Department of Physics, Graduate School of Science, Kyoto University, 
   Kitashirakawa Oiwake-cho, Sakyo-ku, Kyoto 606-8502, Japan}
\altaffiltext{3}{%
   Department of Applied Physics and Electronic Engineering, Faculty of Engineering, 
   University of Miyazaki, \\1-1 Gakuen Kibanadai-Nishi, Miyazaki 889-2192, Japan}
\altaffiltext{4}{%
   Institute of Space and Astronautical Science, 
   3-1-1, Yoshinodai, Sagamihara, Kanagawa 229-8510, Japan}

\email{kamitsukasa@ess.sci.osaka-u.ac.jp}


\maketitle

\begin{abstract}

        We report on the Suzaku results of the mixed-morphology supernova remnant (SNR) G290.1$-$0.8 (MSH\,11-61A). The SNR has an asymmetric structure extended to the southeast and the northwest. In the X-ray spectra of the center and the northwest regions, we discover recombining plasma features with the strong Si Ly$\alpha$ and radiative recombination continuum at $\sim$ 2.7\,keV. These features are the most significant in the northwest region, and the spectra are well-reproduced with a recombining plasma of $kT_{\rm e} = 0.5$\,keV. Whereas the spectra of other regions are expressed by an ionizing plasma of $kT_{\rm e} = 0.6$\,keV. The recombining plasma has over-solar abundances, while the ionizing plasma has roughly solar abundances. Hence they are likely ejecta  and interstellar medium (ISM) origin, respectively. The recombining plasma in the northwest of G290.1$-$0.8 would be generated by a break-out of the supernova ejecta from a high density circumstellar medium to a low density ISM.

\end{abstract}

\section{Introduction}

    A quarter of the X-ray-detected supernova remnants (SNRs) in our Galaxy belong to a mixed-morphology (MM) SNR (\cite{Jones1998}). They have center-filled thermal X-ray emissions in a synchrotron radio shell (\cite{Rho1998}). They are generally associated with molecular clouds, and often contain OH (1720\,MHz) masers {(e.g., \cite{Yusef-Zadeh2003}). Some MM SNRs exhibit the TeV/GeV gamma-ray emissions, indicating the presence of high energy protons interacting with molecular clouds (e.g., \cite{Abdo2010IC443}). These facts suggest that MM SNRs are located in the dense environments which contain massive stars. Therefore, we expect that they are originated in core-collapse supernovae (CC SNe).

    Recently, the Suzaku satellite has discovered strong radiative recombination continua (RRC) in the X-ray spectra of several MM SNRs (e.g., W49B: \cite{Ozawa2009}; IC\,443: \cite{Yamaguchi2009}, \cite{Ohnishi2014}; G359.1$-$0.5: \cite{Ohnishi2011}; W28: \cite{Sawada2012}; W44: \cite{Uchida2012}; G346.6$-$0.2: \cite{Yamauchi2013}; 3C\,391: \cite{Ergin2014}). In the case of W49B, with the XMM-Newton, \citet{Miceli2010} confirmed the RRC feature and found that it is spatially localized. The strong RRC emissions are the direct evidence that these SNRs have recombining plasmas (RPs). The RP is also characterized with a higher ionization temperature, $kT_{\rm z}$, than an electron temperature, $kT_{\rm e}$, unlike a collisional ionization equilibrium (CIE: $kT_{\rm e} = kT_{\rm z}$) or an ionizing plasma (IP: $kT_{\rm e} > kT_{\rm z}$). Although the formation process of the RP has not been understood yet, the strong correlation between the RPs and the MM SNRs suggests that the RP would be formed under the environments for MM SNRs.

    \citet{Kesteven1968} discovered G290.1$-$0.8, also known as MSH\,11-61A, in the radio band. The radio image obtained with the Molonglo Observation Synthesis Telescope (MOST) revealed its asymmetric geometry elongated in the southeast and the northwest direction with the size of \timeform{15'}\,$\times$\,\timeform{10'} (\cite{Kesteven1987}; \cite{Milne1989}).  \citet{Filipovic2005} reported that the NANTEN CO data suggest a dense molecular cloud in the southeast of the SNR. \citet{Reynoso2006} studied the gas distribution and kinematics in detail, and suggested that the SNR probably lies in the Carina arm, at a distance of $7\pm1$\,kpc.

    In X-ray, G290.1$-$0.8 was found by the Galactic SNRs survey with the Einstein Observatory (\cite{Seward1990}). The Advanced Satellite for Cosmology and Astrophysics (ASCA) found that the X-ray is a thermal plasma emission, and then classified this SNR as a member of the MM\,SNRs (\cite{Rho1998}). 
\citet{Garcia2012} analyzed the XMM-Newton and the Chandra data focusing on the asymmetric geometry of the SNR. They concluded that the plasmas in the southeast and the northwest regions are IPs, while those in the other regions are CIEs. Employing these X-ray results and the H$_{\rm I}$ map, they proposed that the SNR is due to a core collapse of a high mass progenitor with a bipolar wind.

    \citet{Pavan2011} discovered a hard X-ray source, IGR J11014-6103, about \timeform{10'} at the southwest of G290.1$-$0.8. \citet{Halpern2014} detected the pulsation of 62.8\,msec. This pulsar has a prominent jet extending toward the northwest and a possible counter jet to the southeast (\cite{Pavan2014}). It also has a bow shock tail pointing to the center of G290.1$-$0.8. Moreover, the absorption ($N_{\rm H}$) of the pulsar is similar to that of G290.1$-$0.8. Thus IGR J11014-6103 would be an associated pulsar of G290.1$-$0.8, a compact remnant of a CC SN. \citet{Halpern2014} estimated the spin-down age to be 116\,kyr, significantly larger than the previous estimation of the SNR age of 10--20 kyr (\cite{Slane2002}). 

    In this paper, we present the Suzaku observation results of G290.1$-$0.8. The size of the SNR allows us to perform a spatially resolved spectroscopic study by using the Suzaku telescope. The high sensitivity and the high energy resolution of Suzaku enable us to accurately determine the X-ray spectral features of the thermal plasma. We report the first discovery of the RP from this SNR.  Based on the detailed analysis  of the plasma structure, we discuss the evolution of G290.1$-$0.8.

\section{Observation and Data Reduction}

    The Suzaku satellite (\cite{Mitsuda2007}) observed G290.1$-$0.8 on 2011 June 25 (PI: K. Koyama) with the X-ray Imaging Spectrometer (XIS; \cite{Koyama2007}). The observation log is given in table \ref{obslog}. The XIS consists of three active sensors placed on the focal planes of the X-ray Telescopes (XRTs; \cite{Serlemitsos2007}). Two of the XIS are front-illuminated (FI) CCDs, sensitive in the 0.4--14\, keV, while the other is a back-illuminated (BI) CCD, with high sensitivity down to 0.2\,keV. The spaced-row charge injection (SCI; \cite{Nakajima2008}) technique was performed for all of the XIS on this observation. We use the version 6.15 of the HEAsoft tools (ver.19 of the Suzaku software) for the data reduction. The archival data are reprocessed with the calibration data base (CALDB) released in 2014 April. The total exposure time after the standard screening\footnote{http://heasarc.nasa.gov/docs/suzaku/processing/criteria\_xis.html} is about 110\,ks. 

\section{Analyses and Results}

\subsection{Imaging Analysis}

    Figure \ref{image} shows X-ray images in the 0.6--1.5 and 1.5--4\,keV bands. We combine the data from all the XIS to maximize the photon statistics. The images are binned with \timeform{2''.1}\,$\times$\,\timeform{2''.1} and smoothed with a Gaussian kernel of $\sigma =$\,\timeform{0'.5}. We overlay the 843\,MHz radio profile obtained by the MOST with white contours. We can see a center-filled X-ray emission in the radio shell, which is the typical feature of the MM SNRs. The diffuse emission from G290.1$-$0.8 is mainly found in the 0.6--4\,keV band; no significant emission except the X-ray background is found above this energy band. We see ear-like structures at the southeast and the northwest that are prominent in the 1.5--4\,keV band. 

\subsection{Spectral Analysis}

    We use the XSPEC software version 12.8.2a in the following analysis. The redistribution matrix files (RMFs) and the ancillary response files (ARFs) are generated with \texttt{xisrmfgen} and \texttt{xisarfgen} (\cite{Ishisaki2007}), respectively. We adopt the solar abundances of \citet{Anders1989}. Unless otherwise specified, all errors represent 90\,\% confidence levels.

\subsubsection{Background Estimation}

    Since  G290.1$-$0.8 extends widely over the field of view (FoV) of the XIS, the X-ray background region is not available from the same FoV. Therefore, we use the data from a nearby sky field: a blank region in the field of the  X-ray pulsar 1E1048.1$-$5937. We extract the background data from the whole FoV of the XIS, excluding the pulsar region of \timeform{4'} radius.  We generate the non X-ray background (NXB) using \texttt{xisnxbgen} (\cite{Tawa2008}) and subtract it from the extracted spectrum. This background spectrum consists of the cosmic X-ray background (CXB), the Galactic ridge X-ray emission (GRXE), and the Galactic halo (GH) (e.g., \cite{Kushino2002}; \cite{Kaneda1997}; \cite{Henley2013}).  Then, we fit the spectrum with a model of [Abs1 $\times$ powerlaw (CXB) + Abs2 $\times$ (apec (HP) + apec (LP)) + Abs3 $\times$ apec (GH)], where the apec is a CIE plasma model in the XSPEC.  The second term is the  GRXE component, which is represented with a 2-CIE model, a high-temperature plasma (HP; $kT_{\rm e}\sim7$\,keV) $+$ a low-temperature plasma (LP; $kT_{\rm e}\sim1$\,keV) (\cite{Kaneda1997}; \cite{Uchiyama2013}).  The CXB component parameters are fixed at those in \citet{Kushino2002}. The model is nicely fitted with $\chi^2_{\nu} / {\rm d.o.f.} = 1.08 /1111$, where $\chi^2_{\nu}$ and ${\rm d.o.f.}$ represent the reduced chi square and the degree of freedom. The best-fit parameters are given in the table \ref{bkg-para}.

    Since the Galactic coordinate of G290.1$-$0.8 is slightly different from that of the 1E1048.1$-$5937 field, we fine-tune the GRXE flux using the spatial structure of the GRXE.  \citet{Kaneda1997} estimated its Galactic latitude distribution by an exponential function with the e-folding values  of  $\sim$ \timeform{0D.5} and \timeform{1D} for the HP and the LP, respectively. \citet{Uchiyama2013} estimated the Galactic longitude distribution by an exponential function with the e-folding values of $\sim$ \timeform{50D} both for the HP and the LP. Then, we can estimate the fluxes of the HP and the LP at G290.1$-$0.8 to be $\sim 60\,\%$ and $\sim 80\,\%$ of those at 1E1048.1$-$5937 field.  

\subsubsection{Analyses of the Spectra}

    Since the spatial structure of G290.1$-$0.8 is asymmetric (\cite{Kesteven1987}), we divide the SNR into the Center, NW, SE, NE, and SW regions as shown in figure \ref{image} (b). The NXB-subtracted spectra are shown in figure \ref{spectra}. We first apply a 1-component IP adding the background model spectrum given in the previous subsection. We use the plasma code (vvrnei) in the XSPEC package. The vvrnei calculates the spectrum of a non-equilibrium ionization plasma after a rapid transition of the electron temperature from $kT_{\rm init}$ to $kT_{\rm e}$. The initial plasma temperature $kT_{\rm init}$ is fixed at 0.01\,keV. The present electron temperature $kT_{\rm e}$, ionization timescale $n_{\rm e}t$, emission measure, and column density $N_{\rm H}$ are free parameters, where $n_{\rm e}$ and $t$ are an electron density and an elapsed time after the initial state. The abundances of Ne, Mg, Si, S, Ar, and Fe are also allowed to vary freely, while the abundances of Ca and Ni are linked to those of Ar and Fe, respectively.  The CXB flux is free within the  fluctuation expected for the area of each region (\cite{Kushino2002}), while the other background parameters are fixed to the values in the table \ref{bkg-para}.

    This fit, however, leaves a large residual around 1.2\,keV. This residual is due to the error of Fe-L lines in the current XSPEC code (e.g., \cite{Borkowski2006}; \cite{Yamaguchi2011}). We therefore add a Gaussian line at 1.2\,keV. The $\chi^2_{\nu}$ values are improved to $1.19$  (Center), $1.24$ (NW), $1.08$ (SE), $1.06$ (NE), and $1.10$ (SW). Contrary to the nice fits for the SE, NE and SW regions, the IP models for the Center and particularly the NW spectra are rejected with the large $\chi^2_{\nu}$ values.  For example, the fits of the NW spectra leave hump-like residuals around 2.7\,keV and line-like residuals at 2.0 keV (see the middle panels of the NW spectra in figure \ref{spectra}). The former residuals correspond to the RRC of Si, while the latters are Si Ly$\alpha$ (2.01\,keV). These residuals indicate that the ionization state of Si is higher than that expected from the IP model. In particular, the RRC structure of Si strongly suggests that the plasma is in a recombining phase.

    We therefore apply RP models for the spectra of the Center and the NW (1-component RP plus a Gaussian line). Unlike the fixed initial temperature $kT_{\rm init}$ of 0.01 keV in the IP fit, the RP fit assumes a high temperature in the initial phase of the plasma. The RP fit assumes a rapid electron cooling of $kT_{\rm init} \rightarrow kT_{\rm e}$ at an early phase of the plasma evolution. Thus the free parameters are $kT_{\rm init}$, $kT_{\rm e}$, and $n_{\rm e}t$, where $t$ is an elapsed time after the rapid cooling of the electron temperature. This RP model fit significantly improves the $\chi^2_{\nu}$ value to 1.12 and 1.10 for the Center and the NW spectra, respectively. The hump-like residuals in the IP fits disappear in the RP fits (see the bottom panels of the Center and NW spectra in figure \ref{spectra}).  On the other hand, the RP fits for the SE, NE and SW spectra do not significantly improve the $\chi^2_{\nu}$ value from those of the IP fits. Although the hump-like residuals also seen in the NE, the current photon statistics does not allow us to discriminate the IP and RP models.

    The best-fit spectra and parameters for the RP fits (Center and NW) and IP fits (SE, NE, and SW) are shown in the figure \ref{spectra} and table \ref{fit-para}. The null hypothesis probabilities of the IP fits are $8\times10^{-7}$ (Center), $9\times10^{-6}$ (NW), $0.03$ (SE), $0.08$ (NE), and $0.01$ (SW), while those of the RP fits are  $0.001$ (Center), $0.03$ (NW). Although all the spectra are statistically rejected under the canonical criterion of the null probability of 0.1, the IP fits for the SE, NE and SW and the RP fits for the Center and NW are marginally acceptable by taking account of possible systematic errors. 
    
    We select the similar regions to those employed in \citet{Garcia2012}. They reported that CIE models adequately describe some of the regions. We therefore try CIE models for the direct comparison between the Suzaku and the XMM-Newton results. The CIE results, however, show no significant difference from the IP fits within the systematical uncertainty; the $\chi^2_{\nu}$ values of the CIE fits are 1.22 (Center), 1.24 (NW), 1.12 (SE), 1.09 (NE), and 1.11 (SW). We thus use the IP fit results for the SE, NE, and SW regions, while the RP results for the Center and NW regions in the following discussion as the good approximation.

\section{Discussion}

    We find that the spectrum of G290.1$-$0.8 is well described by RPs in the Center and NW regions, while those in the SE, NE and SW regions are IPs. The electron temperatures are $0.45_{-0.01}^{+0.02}$\,keV\,(Center), $0.52_{-0.04}^{+0.02}$\,keV (NW), $0.66_{-0.02}^{+0.03}$\,keV (SE), $0.64_{-0.01}^{+0.02}$\,keV (NE), and $0.59_{-0.02}^{+0.03}$\,keV (SW). Thus the temperatures of the RP regions (Center and NW) are lower than those of the IP regions (SE, NE, and SW) about 25\,\%. The ionization parameter ($n_{\rm e}t$) in the NW region is smaller than that in the Center region, which is consistent with the fact that the apparent RRC feature in the NW spectrum is clearer than that in the Center. In contrast, the XMM-Newton observations reported the plasma of IP or CIE in all the regions (\cite{Garcia2012}).  We believe that the discrepancy is due to the  better  energy resolution and  photon statistics in the hard energy band of Suzaku than those of XMM-Newton (as in the case of IC\,443; \cite{Yamaguchi2009}). 

    As we noted in the section 3.2.2, we allowed the possible fluctuation of the CXB in the source regions. We here examine the effect of the CXB fluctuation changing the flux by $\pm 50$\,\%, which is the largest deviation expected for each region, of the nominal value (\cite{Kushino2002}). The results, however, show no significant change of the $\chi^2_{\nu}$ and the best-fit parameters; for example the electron temperatures change only by $0.01$\,keV, which is within the range of the statistical errors.

    The abundances of the RP are over-solar, while those of the IP except in the SE region are roughly 1 solar. The abundances in the SE region seem intermediate between those in the RP and IP regions. Probably, the plasma in the SE region is a mixture of the RP and the IP. In fact, although the significance level is low, the RP fit for the SE ($\chi^2 / {\rm d.o.f.} = 1210 / 1125$) is slightly better than the pure IP fit ($\chi^2 / {\rm d.o.f.} = 1215 / 1126$).
    
    The abundance pattern of the RP in NW is shown in figure \ref{abund_pattern} together with the prediction from the massive progenitor (\cite{Woosley1995}). The pattern is roughly similar to that of ejecta from the progenitor star of $M = 20$--$25\,{M_{\odot}}$. On the other hand, the solar abundances of the IP in the NE and SW regions suggest that the plasma is ISM origin.
    
    In detail, the Ne and Fe abundances in every region are systematically lower than the other elements. The Fe abundances are mainly determined by the line fluxes of Fe-L lines, which have rather large uncertainty in the present plasma code. The main transitions are 3s$\rightarrow$2p ($\sim 0.7$\,keV) and 3d$\rightarrow$2p ($\sim 0.8$\,keV) of Fe\emissiontype{XVII}. These low energy line fluxes are significantly affected by the low energy absorptions and contamination from the Ly$\beta$ of O\emissiontype{VIII} (0.77\,keV). As the results, the Fe abundance may have a large error, and hence apparent depletion of Fe may occur. The Ne abundances are due to K$\alpha$ lines of Ne\emissiontype{IX} and Ne\emissiontype{X}. The L-line energies from more highly ionized Fe\emissiontype{XIX} come near the K$\alpha$ of Ne\emissiontype{IX} (e.g., see \cite{Brinkman2000}). Therefore, coupled to the large error of Fe abundance, the Ne abundance would have a large error too. However, the reason of apparent Ne depletion is somehow puzzling. We simply note that the same depletions of Ne and Fe are observed in \citet{Garcia2012}, and also reported from other SNRs (e.g., W28, \cite{Sawada2012}).

    To form the RP in the Center and the NW regions, either a rapid electron cooling or an enhancement of ionization should occur. In the former case, the electron temperature is expected to be lower in the RP region, while visa versa for the latter case. Since our results show a lower temperature of the RP than that of the IP, the former scenario is preferable; the electron cooling occurred to the center-northwest direction.

    The electron cooling may occur by a thermal conduction from cold clouds (\cite{Cox1999}; \cite{Shelton1999}). In this case, the RP should be prominent in the contact region with cold dense clouds.  \citet{Filipovic2005} reported that a dense molecular cloud may be interacting with the SNR shock at the southwest. However, no hint of interaction such as maser sources is found in the cloud. Furthermore, the position of the molecular cloud is significantly off-set from the RP region (Center-NW).  Moreover, we see no hint of a thermal conduction: no temperature decrease toward the molecular clouds. Thus the thermal conduction is unlikely for  the RP in the Center to the NW of G290.1$-$0.8.

    We therefore discuss an alternative scenario of the rapid electron cooling, an adiabatic rarefaction (\cite{Itoh1989}; \cite{Shimizu2012}). Since the ejecta of RP is aligned from the Center to the NW, and the electron temperature decreases toward the same direction, the rarefaction would be jet-like pointing to the NW. Also the X-ray/radio morphologies of G290.1$-$0.8 extend further to the NW.
    
    Since the best-fit $n_{\rm e} t$ and $kT_{\rm e}$ values may suggest the higher pressure in the RP than that in the IP region, one may argue that this is inconsistent with the rarefaction scenario. However, our rarefaction scenario is not a break-out from the IP (NE and SW) to the RP (Center and NW) regions. We assume that the ejecta became a RP by an adiabatic break-out from the high density circumstellar region (very near the SNR center) to the low density ISM region, while the IP is the result of independent process: canonical ionizing process.

    Assuming the distance of 7\,kpc and using the best-fit $n_{\rm e}t$ and $EM$ in the NW region, the recombining time ($t_{\rm rec}$) is estimated to be $50\,(f/0.25)^{0.5}\,{\rm kyr}$. While this value is larger than the estimated SNR age of 10--20\,kyr (\cite{Slane2002}), This disagreement is probably due to the simplified age estimation of the complex SNR. In fact, Halpern et al. (2014) reported the spin-down age of the pulsar to be 116\,kyr, which favors our recombining time. If the SNR age is equal to the recombining time of $50\,(f/0.25)^{0.5}\,{\rm kyr}$, the kick velocity of the pulsar is $400\,(f/0.25)^{-0.5}\,{\rm km\,s^{-1}}$. This is consistent with the mean value of pulsars of $450\pm90\,{\rm km s^{-1}}$ (\cite{Lyne1994}), and contradicts the high kick velocity of 1000--2000\,${\rm km s^{-1}}$, estimated by \citet{Pavan2014}. 
      
    Since the long-jet of IGR J11014-6103 is pointing to the northwest, the same direction of the RP region, it may be conceivable that the RP was made by the jet when the pulsar was still in the main body of the SNR. However, this scenario would be a remote possibility because the jet energy is far less to make a significant over-ionization at the NW plasma, unless it was extraordinary bright, like a gamma-ray burst and its afterglow.

\section{Summary}

    We have analyzed Suzaku/XIS data obtained from G290.1$-$0.8. The results are summarized as follows:

\begin{enumerate}

\item    The plasma states in G290.1$-$0.8 are different from region to region. We find Si Ly$\alpha$ and RRCs from the NW and the Center sepctra, while not from other regions. Thus the plasmas are in the recombining phase at the Center and the NW, while in the ionizing phase at other regions. 

\item    The electron temperature of the RP is lower than that of the IP.

\item    The abundance pattern indicates that the RP is dominated by ejecta of CC SN, while the IP is likely ISM origin.

\item    A plausible origin of the RP is the adiabatic rarefaction to the NW direction.

\end{enumerate}

\section*{Acknowledgment}

    We thank all members of the Suzaku operation and calibration teams. This work is supported by Japan Society for the Promotion of Science (JSPS) KAKENHI Grant Number 23000004, 23340071, 24540229, 24684010,  24740167, 25800119, 26800102. 

\newpage
\bibliographystyle{pasj}
\bibliography{reference}

\begin{thebibliography}{}
\expandafter\ifx\csname natexlab\endcsname\relax\def\natexlab#1{#1}\fi

\bibitem[{{Abdo} {et~al.}(2010){Abdo}, {Ackermann}, {Ajello}, {Baldini},
  {Ballet}, {Barbiellini}, {Bastieri}, {Baughman}, {Bechtol}, {Bellazzini},
  {Berenji}, {Blandford}, {Bloom}, {Bonamente}, {Borgland}, {Bregeon}, {Brez},
  {Brigida}, {Bruel}, {Burnett}, {Buson}, {Caliandro}, {Cameron}, {Caraveo},
  {Casandjian}, {Cecchi}, {{\c C}elik}, {Chekhtman}, {Cheung}, {Chiang},
  {Cillis}, {Ciprini}, {Claus}, {Cohen-Tanugi}, {Cominsky}, {Conrad}, {Cutini},
  {Dermer}, {de Angelis}, {de Palma}, {Silva}, {Drell}, {Drlica-Wagner},
  {Dubois}, {Dumora}, {Farnier}, {Favuzzi}, {Fegan}, {Focke}, {Fortin},
  {Frailis}, {Fukazawa}, {Funk}, {Fusco}, {Gargano}, {Gasparrini}, {Gehrels},
  {Germani}, {Giavitto}, {Giebels}, {Giglietto}, {Giordano}, {Glanzman},
  {Godfrey}, {Grenier}, {Grondin}, {Grove}, {Guillemot}, {Guiriec}, {Hanabata},
  {Harding}, {Hayashida}, {Hughes}, {Jackson}, {J{\'o}hannesson}, {Johnson},
  {Johnson}, {Johnson}, {Kamae}, {Katagiri}, {Kataoka}, {Kawai}, {Kerr},
  {Kn{\"o}dlseder}, {Kocian}, {Kuss}, {Lande}, {Latronico}, {Lee},
  {Lemoine-Goumard}, {Longo}, {Loparco}, {Lott}, {Lovellette}, {Lubrano},
  {Madejski}, {Makeev}, {Mazziotta}, {Meurer}, {Michelson}, {Mitthumsiri},
  {Moiseev}, {Monte}, {Monzani}, {Morselli}, {Moskalenko}, {Murgia},
  {Nakamori}, {Nolan}, {Norris}, {Nuss}, {Ohsugi}, {Orlando}, {Ormes}, {Ozaki},
  {Paneque}, {Panetta}, {Parent}, {Pelassa}, {Pepe}, {Pesce-Rollins}, {Piron},
  {Porter}, {Rain{\`o}}, {Rando}, {Razzano}, {Reimer}, {Reimer}, {Reposeur},
  {Rochester}, {Rodriguez}, {Romani}, {Roth}, {Ryde}, {Sadrozinski}, {Sanchez},
  {Sander}, {Saz Parkinson}, {Scargle}, {Sgr{\`o}}, {Siskind}, {Smith},
  {Smith}, {Spandre}, {Spinelli}, {Strickman}, {Strong}, {Suson}, {Tajima},
  {Takahashi}, {Takahashi}, {Tanaka}, {Thayer}, {Thayer}, {Thompson},
  {Tibaldo}, {Torres}, {Tosti}, {Tramacere}, {Uchiyama}, {Usher}, {Van Etten},
  {Vasileiou}, {Venter}, {Vilchez}, {Vitale}, {Waite}, {Wang}, {Winer}, {Wood},
  {Ylinen}, \& {Ziegler}}]{Abdo2010IC443}
{Abdo}, A.~A., {et~al.} 2010, \apj, 712, 459

\bibitem[{{Anders} \& {Grevesse}(1989)}]{Anders1989}
{Anders}, E., \& {Grevesse}, N. 1989, \gca, 53, 197

\bibitem[{{Borkowski} {et~al.}(2006){Borkowski}, {Hendrick}, \&
  {Reynolds}}]{Borkowski2006}
{Borkowski}, K.~J., {Hendrick}, S.~P., \& {Reynolds}, S.~P. 2006, \apj, 652,
  1259

\bibitem[{{Brinkman} {et~al.}(2000){Brinkman}, {Gunsing}, {Kaastra}, {van der
  Meer}, {Mewe}, {Paerels}, {Raassen}, {van Rooijen}, {Br{\"a}uninger},
  {Burkert}, {Burwitz}, {Hartner}, {Predehl}, {Ness}, {Schmitt}, {Drake},
  {Johnson}, {Juda}, {Kashyap}, {Murray}, {Pease}, {Ratzlaff}, \&
  {Wargelin}}]{Brinkman2000}
{Brinkman}, A.~C., {et~al.} 2000, \apjl, 530, L111

\bibitem[{{Cox} {et~al.}(1999){Cox}, {Shelton}, {Maciejewski}, {Smith},
  {Plewa}, {Pawl}, \& {R{\'o}{\.z}yczka}}]{Cox1999}
{Cox}, D.~P., {Shelton}, R.~L., {Maciejewski}, W., {Smith}, R.~K., {Plewa}, T.,
  {Pawl}, A., \& {R{\'o}{\.z}yczka}, M. 1999, \apj, 524, 179

\bibitem[{{Ergin} {et~al.}(2014){Ergin}, {Sezer}, {Saha}, {Majumdar},
  {Chatterjee}, {Bayirli}, \& {Ercan}}]{Ergin2014}
{Ergin}, T., {Sezer}, A., {Saha}, L., {Majumdar}, P., {Chatterjee}, A.,
  {Bayirli}, A., \& {Ercan}, E.~N. 2014, \apj, 790, 65

\bibitem[{{Filipovic} {et~al.}(2005){Filipovic}, {Payne}, \&
  {Jones}}]{Filipovic2005}
{Filipovic}, M.~D., {Payne}, J.~L., \& {Jones}, P.~A. 2005, Serbian
  Astronomical Journal, 170, 47

\bibitem[{{Garc{\'{\i}}a} {et~al.}(2012){Garc{\'{\i}}a}, {Combi},
  {Albacete-Colombo}, {Romero}, {Bocchino}, \&
  {L{\'o}pez-Santiago}}]{Garcia2012}
{Garc{\'{\i}}a}, F., {Combi}, J.~A., {Albacete-Colombo}, J.~F., {Romero},
  G.~E., {Bocchino}, F., \& {L{\'o}pez-Santiago}, J. 2012, \aap, 546, A91

\bibitem[{{Halpern} {et~al.}(2014){Halpern}, {Tomsick}, {Gotthelf}, {Camilo},
  {Ng}, {Bodaghee}, {Rodriguez}, {Chaty}, \& {Rahoui}}]{Halpern2014}
{Halpern}, J.~P., {et~al.} 2014, \apjl, 795, L27

\bibitem[{{Henley} \& {Shelton}(2013)}]{Henley2013}
{Henley}, D.~B., \& {Shelton}, R.~L. 2013, \apj, 773, 92

\bibitem[{{Ishisaki} {et~al.}(2007){Ishisaki}, {Maeda}, {Fujimoto}, {Ozaki},
  {Ebisawa}, {Takahashi}, {Ueda}, {Ogasaka}, {Ptak}, {Mukai}, {Hamaguchi},
  {Hirayama}, {Kotani}, {Kubo}, {Shibata}, {Ebara}, {Furuzawa}, {Iizuka},
  {Inoue}, {Mori}, {Okada}, {Yokoyama}, {Matsumoto}, {Nakajima}, {Yamaguchi},
  {Anabuki}, {Tawa}, {Nagai}, {Katsuda}, {Hayashida}, {Bamba}, {Miller},
  {Sato}, \& {Yamasaki}}]{Ishisaki2007}
{Ishisaki}, Y., {et~al.} 2007, \pasj, 59, 113

\bibitem[{{Itoh} \& {Masai}(1989)}]{Itoh1989}
{Itoh}, H., \& {Masai}, K. 1989, \mnras, 236, 885

\bibitem[{{Jones} {et~al.}(1998){Jones}, {Rudnick}, {Jun}, {Borkowski},
  {Dubner}, {Frail}, {Kang}, {Kassim}, \& {McCray}}]{Jones1998}
{Jones}, T.~W., {et~al.} 1998, \pasp, 110, 125

\bibitem[{{Kaneda} {et~al.}(1997){Kaneda}, {Makishima}, {Yamauchi}, {Koyama},
  {Matsuzaki}, \& {Yamasaki}}]{Kaneda1997}
{Kaneda}, H., {Makishima}, K., {Yamauchi}, S., {Koyama}, K., {Matsuzaki}, K.,
  \& {Yamasaki}, N.~Y. 1997, \apj, 491, 638

\bibitem[{{Kesteven} \& {Caswell}(1987)}]{Kesteven1987}
{Kesteven}, M.~J., \& {Caswell}, J.~L. 1987, \aap, 183, 118

\bibitem[{{Kesteven}(1968)}]{Kesteven1968}
{Kesteven}, M.~J.~L. 1968, Australian Journal of Physics, 21, 739

\bibitem[{{Koyama} {et~al.}(2007){Koyama}, {Tsunemi}, {Dotani}, {Bautz},
  {Hayashida}, {Tsuru}, {Matsumoto}, {Ogawara}, {Ricker}, {Doty}, {Kissel},
  {Foster}, {Nakajima}, {Yamaguchi}, {Mori}, {Sakano}, {Hamaguchi},
  {Nishiuchi}, {Miyata}, {Torii}, {Namiki}, {Katsuda}, {Matsuura}, {Miyauchi},
  {Anabuki}, {Tawa}, {Ozaki}, {Murakami}, {Maeda}, {Ichikawa}, {Prigozhin},
  {Boughan}, {Lamarr}, {Miller}, {Burke}, {Gregory}, {Pillsbury}, {Bamba},
  {Hiraga}, {Senda}, {Katayama}, {Kitamoto}, {Tsujimoto}, {Kohmura}, {Tsuboi},
  \& {Awaki}}]{Koyama2007}
{Koyama}, K., {et~al.} 2007, \pasj, 59, 23

\bibitem[{{Kushino} {et~al.}(2002){Kushino}, {Ishisaki}, {Morita}, {Yamasaki},
  {Ishida}, {Ohashi}, \& {Ueda}}]{Kushino2002}
{Kushino}, A., {Ishisaki}, Y., {Morita}, U., {Yamasaki}, N.~Y., {Ishida}, M.,
  {Ohashi}, T., \& {Ueda}, Y. 2002, \pasj, 54, 327

\bibitem[{{Lyne} \& {Lorimer}(1994)}]{Lyne1994}
{Lyne}, A.~G., \& {Lorimer}, D.~R. 1994, \nat, 369, 127

\bibitem[{{Miceli} {et~al.}(2010){Miceli}, {Bocchino}, {Decourchelle},
  {Ballet}, \& {Reale}}]{Miceli2010}
{Miceli}, M., {Bocchino}, F., {Decourchelle}, A., {Ballet}, J., \& {Reale}, F.
  2010, \aap, 514, L2

\bibitem[{{Milne} {et~al.}(1989){Milne}, {Caswell}, {Kesteven}, {Haynes}, \&
  {Roger}}]{Milne1989}
{Milne}, D.~K., {Caswell}, J.~L., {Kesteven}, M.~J., {Haynes}, R.~F., \&
  {Roger}, R.~S. 1989, Proceedings of the Astronomical Society of Australia, 8,
  187

\bibitem[{{Mitsuda} {et~al.}(2007){Mitsuda}, {Bautz}, {Inoue}, {Kelley},
  {Koyama}, {Kunieda}, {Makishima}, {Ogawara}, {Petre}, {Takahashi}, {Tsunemi},
  {White}, {Anabuki}, {Angelini}, {Arnaud}, {Awaki}, {Bamba}, {Boyce}, {Brown},
  {Chan}, {Cottam}, {Dotani}, {Doty}, {Ebisawa}, {Ezoe}, {Fabian}, {Figueroa},
  {Fujimoto}, {Fukazawa}, {Furusho}, {Furuzawa}, {Gendreau}, {Griffiths},
  {Haba}, {Hamaguchi}, {Harrus}, {Hasinger}, {Hatsukade}, {Hayashida}, {Henry},
  {Hiraga}, {Holt}, {Hornschemeier}, {Hughes}, {Hwang}, {Ishida}, {Ishisaki},
  {Isobe}, {Itoh}, {Iyomoto}, {Kahn}, {Kamae}, {Katagiri}, {Kataoka},
  {Katayama}, {Kawai}, {Kilbourne}, {Kinugasa}, {Kissel}, {Kitamoto}, {Kohama},
  {Kohmura}, {Kokubun}, {Kotani}, {Kotoku}, {Kubota}, {Madejski}, {Maeda},
  {Makino}, {Markowitz}, {Matsumoto}, {Matsumoto}, {Matsuoka}, {Matsushita},
  {McCammon}, {Mihara}, {Misaki}, {Miyata}, {Mizuno}, {Mori}, {Mori}, {Morii},
  {Moseley}, {Mukai}, {Murakami}, {Murakami}, {Mushotzky}, {Nagase}, {Namiki},
  {Negoro}, {Nakazawa}, {Nousek}, {Okajima}, {Ogasaka}, {Ohashi}, {Oshima},
  {Ota}, {Ozaki}, {Ozawa}, {Parmar}, {Pence}, {Porter}, {Reeves}, {Ricker},
  {Sakurai}, {Sanders}, {Senda}, {Serlemitsos}, {Shibata}, {Soong}, {Smith},
  {Suzuki}, {Szymkowiak}, {Takahashi}, {Tamagawa}, {Tamura}, {Tamura},
  {Tanaka}, {Tashiro}, {Tawara}, {Terada}, {Terashima}, {Tomida}, {Torii},
  {Tsuboi}, {Tsujimoto}, {Tsuru}, {Turner}, {Ueda}, {Ueno}, {Ueno}, {Uno},
  {Urata}, {Watanabe}, {Yamamoto}, {Yamaoka}, {Yamasaki}, {Yamashita},
  {Yamauchi}, {Yamauchi}, {Yaqoob}, {Yonetoku}, \& {Yoshida}}]{Mitsuda2007}
{Mitsuda}, K., {et~al.} 2007, \pasj, 59, 1

\bibitem[{{Nakajima} {et~al.}(2008){Nakajima}, {Yamaguchi}, {Matsumoto},
  {Tsuru}, {Koyama}, {Tsunemi}, {Hayashida}, {Torii}, {Namiki}, {Katsuda},
  {Shoji}, {Matsuura}, {Miyauchi}, {Dotani}, {Ozaki}, {Murakami}, {Bautz},
  {Kissel}, {Lamarr}, \& {Prigozhin}}]{Nakajima2008}
{Nakajima}, H., {et~al.} 2008, \pasj, 60, 1

\bibitem[{{Ohnishi} {et~al.}(2011){Ohnishi}, {Koyama}, {Tsuru}, {Masai},
  {Yamaguchi}, \& {Ozawa}}]{Ohnishi2011}
{Ohnishi}, T., {Koyama}, K., {Tsuru}, T.~G., {Masai}, K., {Yamaguchi}, H., \&
  {Ozawa}, M. 2011, \pasj, 63, 527

\bibitem[{{Ohnishi} {et~al.}(2014){Ohnishi}, {Uchida}, {Tsuru}, {Koyama},
  {Masai}, \& {Sawada}}]{Ohnishi2014}
{Ohnishi}, T., {Uchida}, H., {Tsuru}, T.~G., {Koyama}, K., {Masai}, K., \&
  {Sawada}, M. 2014, \apj, 784, 74

\bibitem[{{Ozawa} {et~al.}(2009){Ozawa}, {Koyama}, {Yamaguchi}, {Masai}, \&
  {Tamagawa}}]{Ozawa2009}
{Ozawa}, M., {Koyama}, K., {Yamaguchi}, H., {Masai}, K., \& {Tamagawa}, T.
  2009, \apjl, 706, L71

\bibitem[{{Pavan} {et~al.}(2011){Pavan}, {Bozzo}, {P{\"u}hlhofer}, {Ferrigno},
  {Balbo}, \& {Walter}}]{Pavan2011}
{Pavan}, L., {Bozzo}, E., {P{\"u}hlhofer}, G., {Ferrigno}, C., {Balbo}, M., \&
  {Walter}, R. 2011, \aap, 533, A74

\bibitem[{{Pavan} {et~al.}(2014){Pavan}, {Bordas}, {P{\"u}hlhofer},
  {Filipovi{\'c}}, {De Horta}, {O'Brien}, {Balbo}, {Walter}, {Bozzo},
  {Ferrigno}, {Crawford}, \& {Stella}}]{Pavan2014}
{Pavan}, L., {et~al.} 2014, \aap, 562, A122

\bibitem[{{Reynoso} {et~al.}(2006){Reynoso}, {Johnston}, {Green}, \&
  {Koribalski}}]{Reynoso2006}
{Reynoso}, E.~M., {Johnston}, S., {Green}, A.~J., \& {Koribalski}, B.~S. 2006,
  \mnras, 369, 416

\bibitem[{{Rho} \& {Petre}(1998)}]{Rho1998}
{Rho}, J., \& {Petre}, R. 1998, \apjl, 503, L167

\bibitem[{{Sawada} \& {Koyama}(2012)}]{Sawada2012}
{Sawada}, M., \& {Koyama}, K. 2012, \pasj, 64, 81

\bibitem[{{Serlemitsos} {et~al.}(2007){Serlemitsos}, {Soong}, {Chan},
  {Okajima}, {Lehan}, {Maeda}, {Itoh}, {Mori}, {Iizuka}, {Itoh}, {Inoue},
  {Okada}, {Yokoyama}, {Itoh}, {Ebara}, {Nakamura}, {Suzuki}, {Ishida},
  {Hayakawa}, {Inoue}, {Okuma}, {Kubota}, {Suzuki}, {Osawa}, {Yamashita},
  {Kunieda}, {Tawara}, {Ogasaka}, {Furuzawa}, {Tamura}, {Shibata}, {Haba},
  {Naitou}, \& {Misaki}}]{Serlemitsos2007}
{Serlemitsos}, P.~J., {et~al.} 2007, \pasj, 59, 9

\bibitem[{{Seward}(1990)}]{Seward1990}
{Seward}, F.~D. 1990, \apjs, 73, 781

\bibitem[{{Shelton} {et~al.}(1999){Shelton}, {Cox}, {Maciejewski}, {Smith},
  {Plewa}, {Pawl}, \& {R{\'o}{\.z}yczka}}]{Shelton1999}
{Shelton}, R.~L., {Cox}, D.~P., {Maciejewski}, W., {Smith}, R.~K., {Plewa}, T.,
  {Pawl}, A., \& {R{\'o}{\.z}yczka}, M. 1999, \apj, 524, 192

\bibitem[{{Shimizu} {et~al.}(2012){Shimizu}, {Masai}, \&
  {Koyama}}]{Shimizu2012}
{Shimizu}, T., {Masai}, K., \& {Koyama}, K. 2012, \pasj, 64, 24

\bibitem[{{Slane} {et~al.}(2002){Slane}, {Smith}, {Hughes}, \&
  {Petre}}]{Slane2002}
{Slane}, P., {Smith}, R.~K., {Hughes}, J.~P., \& {Petre}, R. 2002, \apj, 564,
  284

\bibitem[{{Tawa} {et~al.}(2008){Tawa}, {Hayashida}, {Nagai}, {Nakamoto},
  {Tsunemi}, {Yamaguchi}, {Ishisaki}, {Miller}, {Mizuno}, {Dotani}, {Ozaki}, \&
  {Katayama}}]{Tawa2008}
{Tawa}, N., {et~al.} 2008, \pasj, 60, 11

\bibitem[{{Uchida} {et~al.}(2012){Uchida}, {Koyama}, {Yamaguchi}, {Sawada},
  {Ohnishi}, {Tsuru}, {Tanaka}, {Yoshiike}, \& {Fukui}}]{Uchida2012}
{Uchida}, H., {et~al.} 2012, \pasj, 64, 141

\bibitem[{{Uchiyama} {et~al.}(2013){Uchiyama}, {Nobukawa}, {Tsuru}, \&
  {Koyama}}]{Uchiyama2013}
{Uchiyama}, H., {Nobukawa}, M., {Tsuru}, T.~G., \& {Koyama}, K. 2013, \pasj,
  65, 19

\bibitem[{{Woosley} \& {Weaver}(1995)}]{Woosley1995}
{Woosley}, S.~E., \& {Weaver}, T.~A. 1995, \apjs, 101, 181

\bibitem[{{Yamaguchi} {et~al.}(2011){Yamaguchi}, {Koyama}, \&
  {Uchida}}]{Yamaguchi2011}
{Yamaguchi}, H., {Koyama}, K., \& {Uchida}, H. 2011, \pasj, 63, 837

\bibitem[{{Yamaguchi} {et~al.}(2009){Yamaguchi}, {Ozawa}, {Koyama}, {Masai},
  {Hiraga}, {Ozaki}, \& {Yonetoku}}]{Yamaguchi2009}
{Yamaguchi}, H., {Ozawa}, M., {Koyama}, K., {Masai}, K., {Hiraga}, J.~S.,
  {Ozaki}, M., \& {Yonetoku}, D. 2009, \apjl, 705, L6

\bibitem[{{Yamauchi} {et~al.}(2013){Yamauchi}, {Nobukawa}, {Koyama}, \&
  {Yonemori}}]{Yamauchi2013}
{Yamauchi}, S., {Nobukawa}, M., {Koyama}, K., \& {Yonemori}, M. 2013, \pasj,
  65, 6

\bibitem[{{Yusef-Zadeh} {et~al.}(2003){Yusef-Zadeh}, {Wardle}, {Rho}, \&
  {Sakano}}]{Yusef-Zadeh2003}
{Yusef-Zadeh}, F., {Wardle}, M., {Rho}, J., \& {Sakano}, M. 2003, \apj, 585,
  319

\end{thebibliography}

\newpage

\begin{table*}
\begin{center}
\caption{Observation logs.}\label{obslog}
\begin{tabular}{lccccc}
\hline
& Target & Obs. ID & Obs. Date & ($l$, $b$) & Exposure\\
\hline
Source &G290.1$-$0.8 & 506061010 & 2011-Jun-25 & (\timeform{290D.1}, \timeform{-0D.75}) & 110.6 ks\\ 
Background & 1E1048.1$-$5937 & 403005010 & 2008-Nov-30 & (\timeform{288D.3}, \timeform{-0D.52}) & 100.4 ks\\ 
\hline
\end{tabular}
\end{center}
\end{table*}

\vspace*{\stretch{4}}
\begin{figure*}
   \begin{center}
         \FigureFile(140mm,140mm){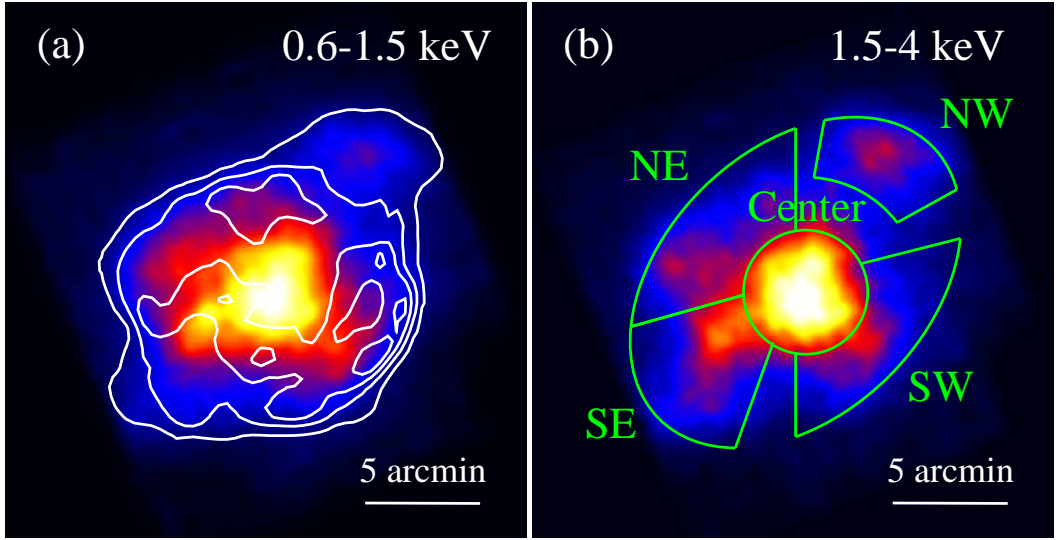}
   \end{center}
   \caption{XIS images of G290.1$-$0.8 in the energy bands of (a) 0.6--1.5\,keV and (b) 1.5--4\,keV, respectively. The 843\,MHz radio profile is overlaid on (a) with white contours. The spectral extraction regions are shown with the green lines in (b).}
\label{image} 
\end{figure*}

\newpage
\begin{table*}
\caption{Best-fit parameters for the background spectrum.}\label{bkg-para}
\begin{center}
\begin{tabular}{llc}
\hline\hline
Component & Parameter & Value\\
\hline
Abs1 & $N_{\rm H}$ $(10^{21} {\rm cm^{-2}})$ & 13.2 (fixed) \\
CXB & photon index & 1.4 (fixed) \\
        & flux$^\dag$ & 1.94 (fixed) \\ \hline
Abs2 & $N_{\rm H}$ $(10^{21} {\rm cm^{-2}})$ & $7.0\pm0.8$ \\
HP & $kT_{\rm e}$ (keV) & 7 (fixed)\\
     & all elements$^\ddag$ & $< 0.2$ \\
LP & $kT_{\rm e}$ (keV) & 1 (fixed)\\
     & all elements$^\ddag$ & $0.10\pm0.02$\\ \hline
Abs3 & $N_{\rm H}$ $(10^{21} {\rm cm^{-2}})$ & $4.1\pm0.7$ \\
GH & $kT_{\rm e}$ (keV) & $0.16\pm0.01$ \\
      & O$^\ddag$ & $0.06_{-0.02}^{+0.03}$ \\
      & Ne$^\ddag$ & $0.17\pm0.02$ \\
      & Mg$^\ddag$ & $0.28_{-0.14}^{+0.15}$ \\
      & Others$^\ddag$ & 1 (fixed) \\
\hline
$\chi^2_{\nu} / {\rm d.o.f.}$ & & $1.08/1111$\\
\hline
\multicolumn{3}{l}{\small $^\dag$ Flux ($10^{-11}\,{\rm erg\,cm^{-2}\,s^{-1}\,deg^{-2}}$) in the $2-10$\,keV band.}\\[-1.5mm]
\multicolumn{3}{l}{\small $^\ddag$ Relative to the solar value (\cite{Anders1989}).}
\end{tabular}
\end{center}
\end{table*}

\newpage
\begin{figure}
	\begin{minipage}{0.5\hsize}
	\begin{center}
		\FigureFile(80mm,80mm){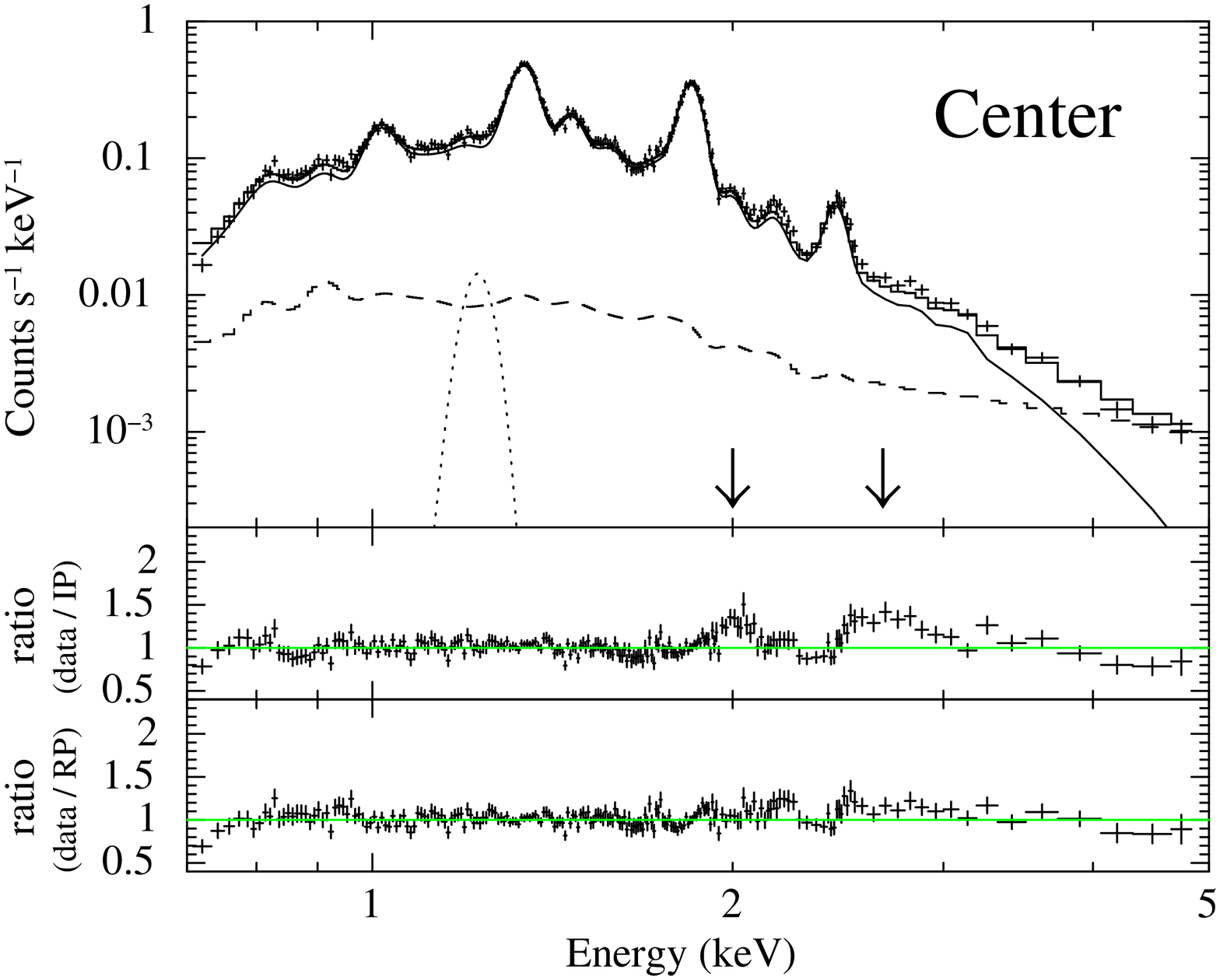}
	\end{center}
	\end{minipage}
	\begin{minipage}{0.5\hsize}
	\begin{center}
		\FigureFile(80mm,80mm){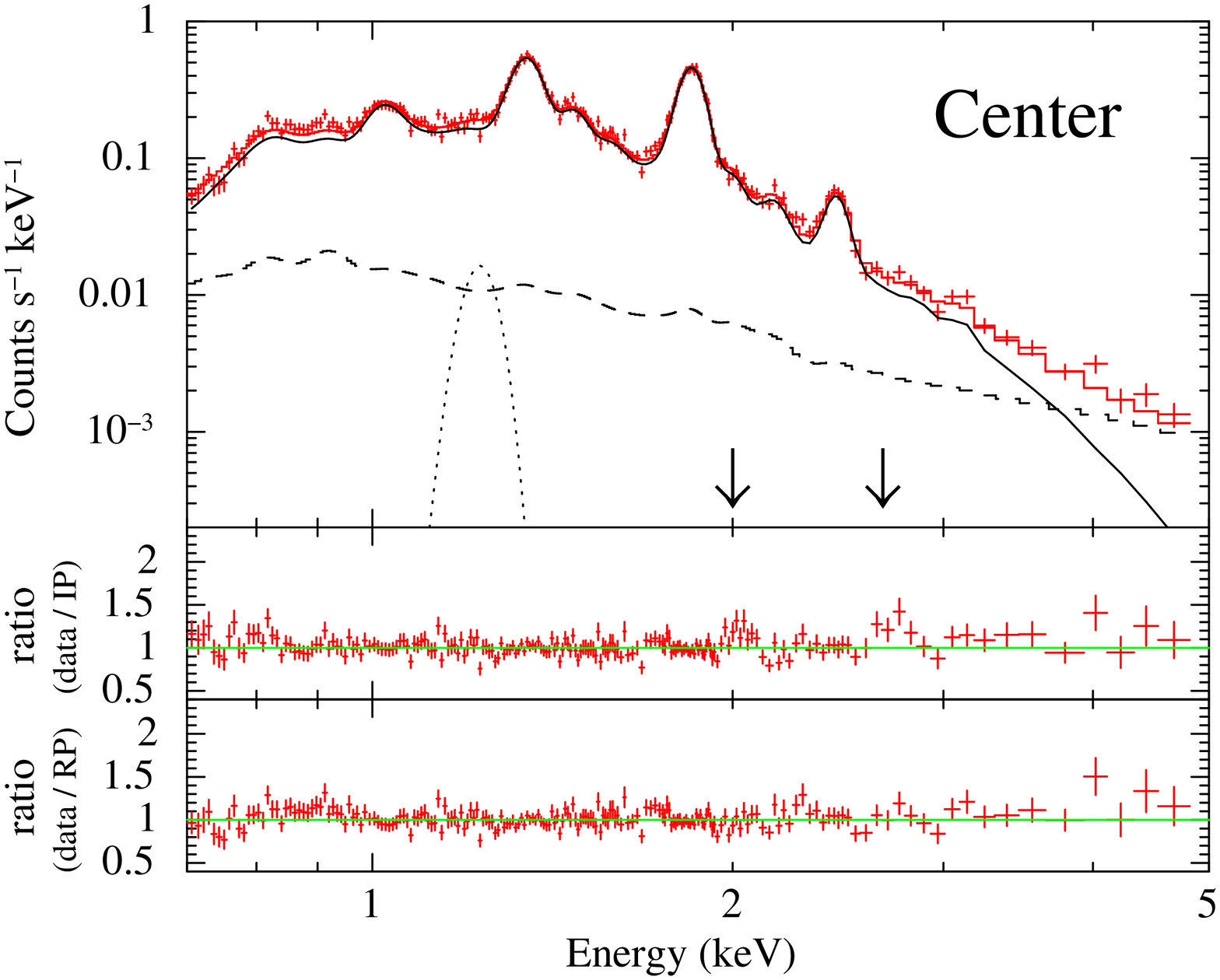}
	\end{center}
	\end{minipage}
	\begin{minipage}{0.5\hsize}
	\begin{center}
		\FigureFile(80mm,80mm){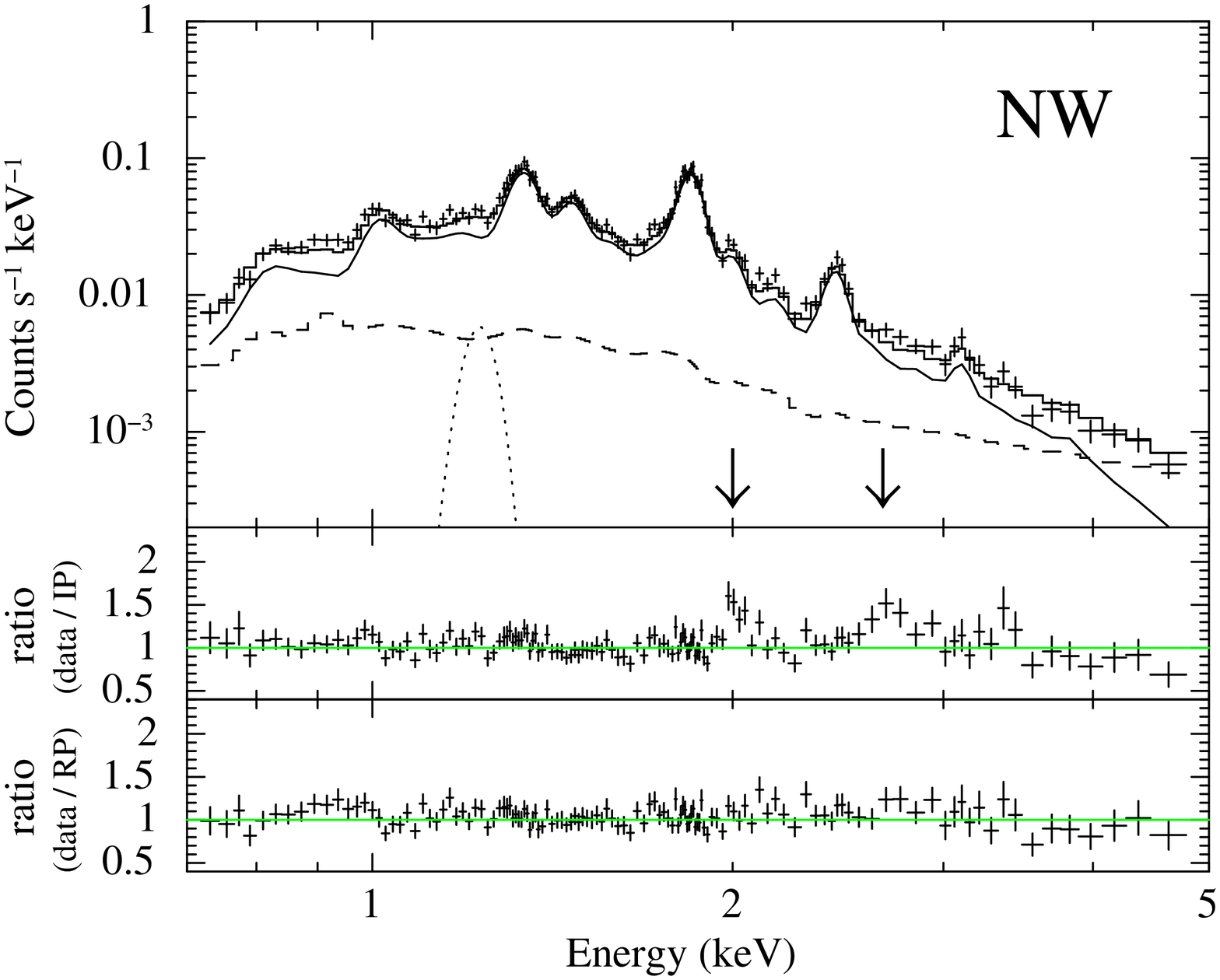}
	\end{center}
	\end{minipage}
	\begin{minipage}{0.5\hsize}
	\begin{center}
		\FigureFile(80mm,80mm){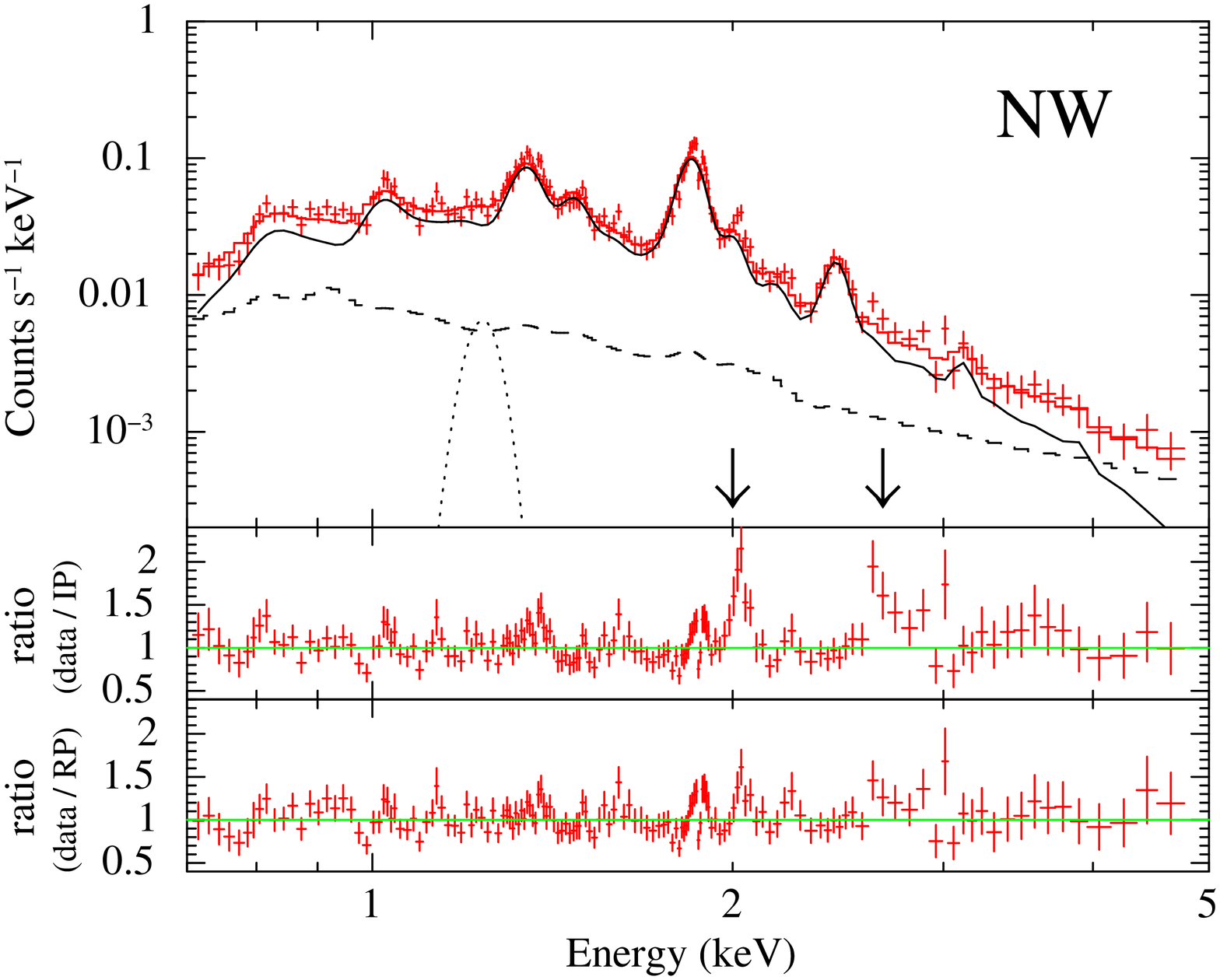}
	\end{center}
	\end{minipage}
	\begin{minipage}{0.5\hsize}
	\begin{center}
		\FigureFile(80mm,80mm){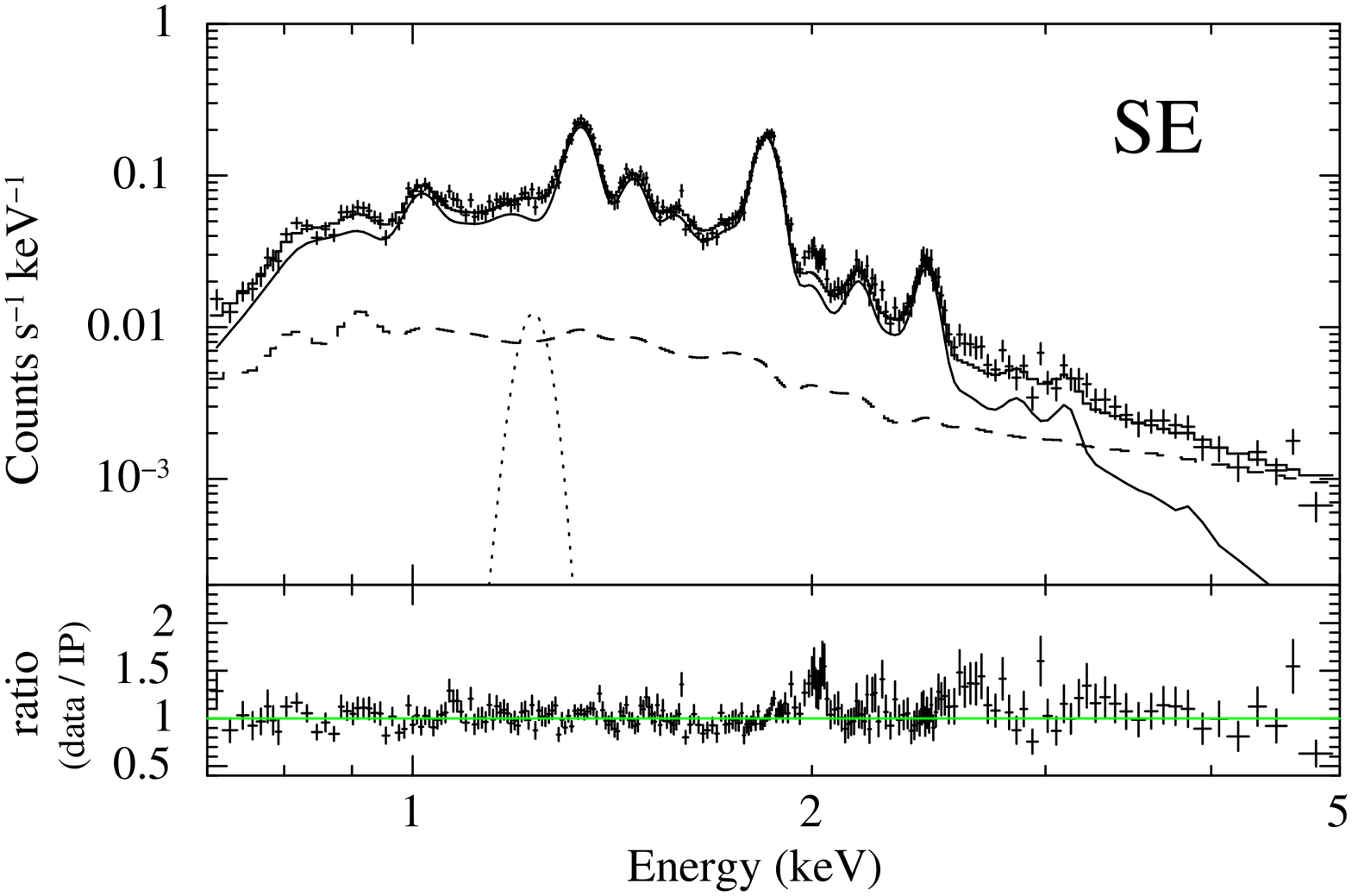}
	\end{center}
	\end{minipage}
	\begin{minipage}{0.5\hsize}
	\begin{center}
		\FigureFile(80mm,80mm){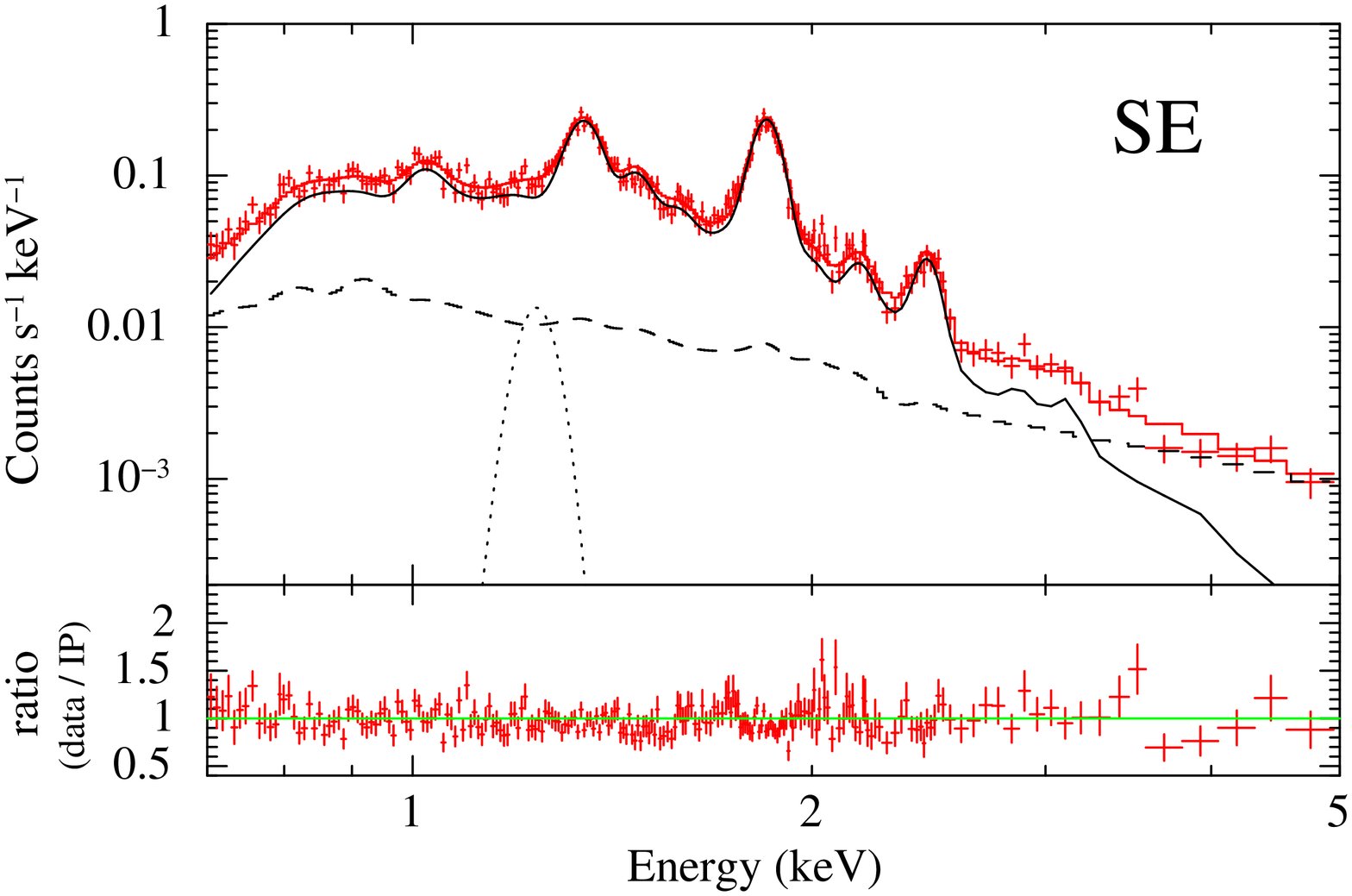}
	\end{center}
	\end{minipage}
	\caption{Spectra obtained from five regions (black: FI, red: BI). Each spectrum is fitted with the plasma model (solid line) + the background model (dashed line) + one Gaussian (dotted line). For the plasma model, the RP is employed in the Center and the NW, while the IP is employed in the SE, NE, and SW. The bottom panels are the ratios of the data to the models we applied. For the Center and the NW, we show the ratios of the data to the IP models in the middle panels for comparison. Error bars represent 1\,$\sigma$ confidence levels.}
\label{spectra}
\end{figure}

\newpage
\setcounter{figure}{1}
\begin{figure}
	\begin{minipage}{0.5\hsize}
	\begin{center}
		\FigureFile(80mm,80mm){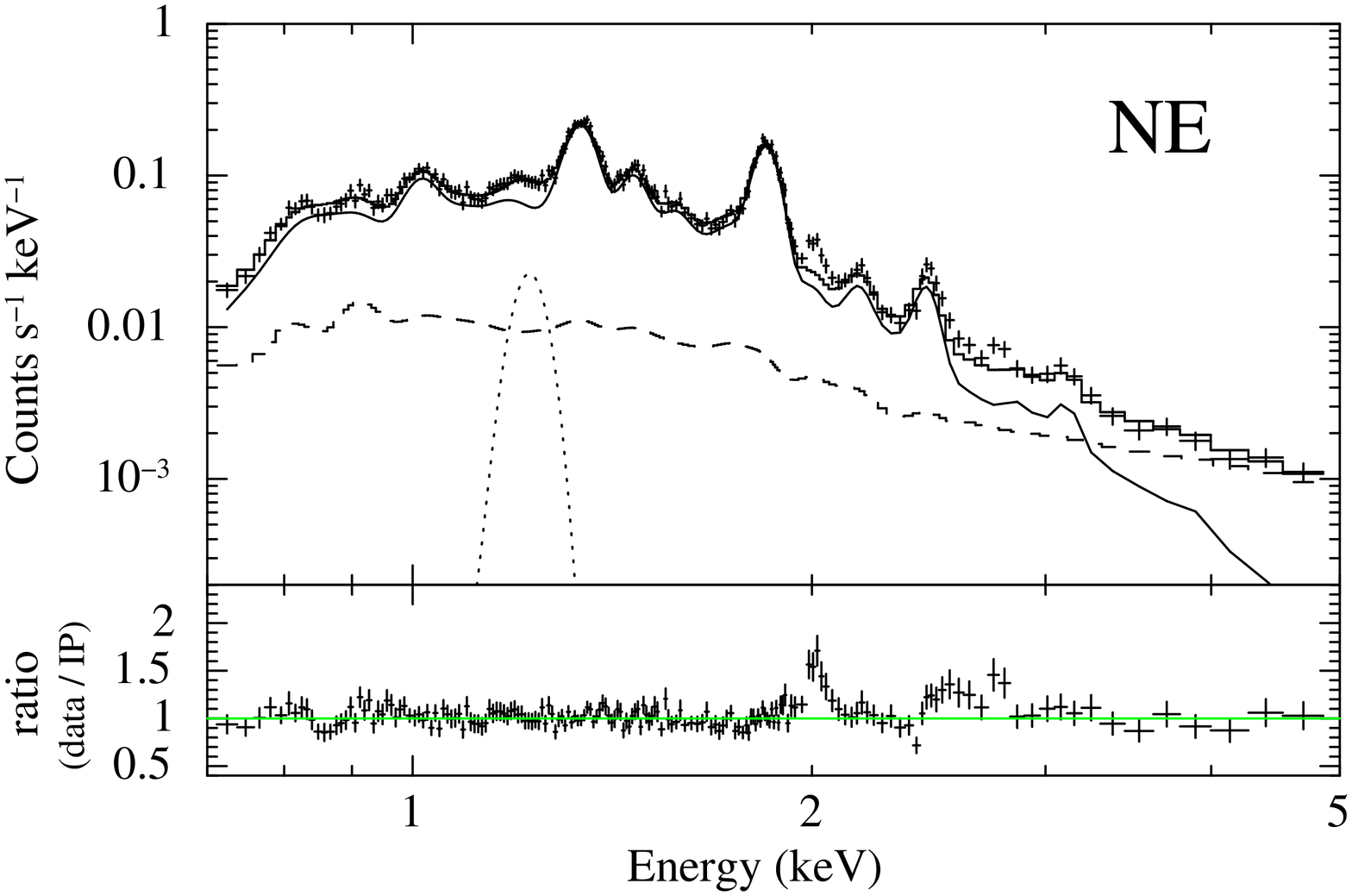}
	\end{center}
	\end{minipage}
	\begin{minipage}{0.5\hsize}
	\begin{center}
		\FigureFile(80mm,80mm){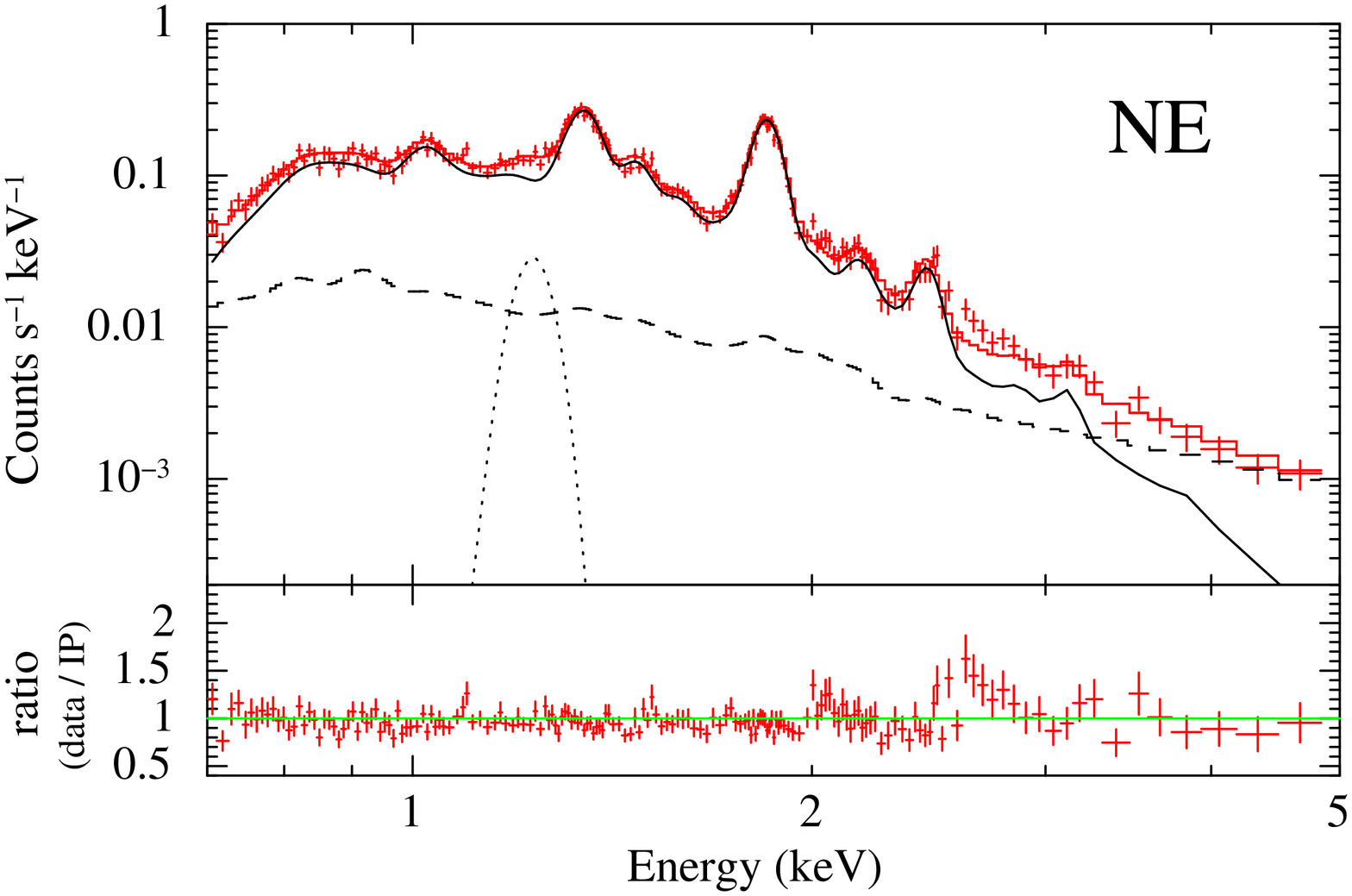}
	\end{center}
	\end{minipage}
	\begin{minipage}{0.5\hsize}
	\begin{center}
		\FigureFile(80mm,80mm){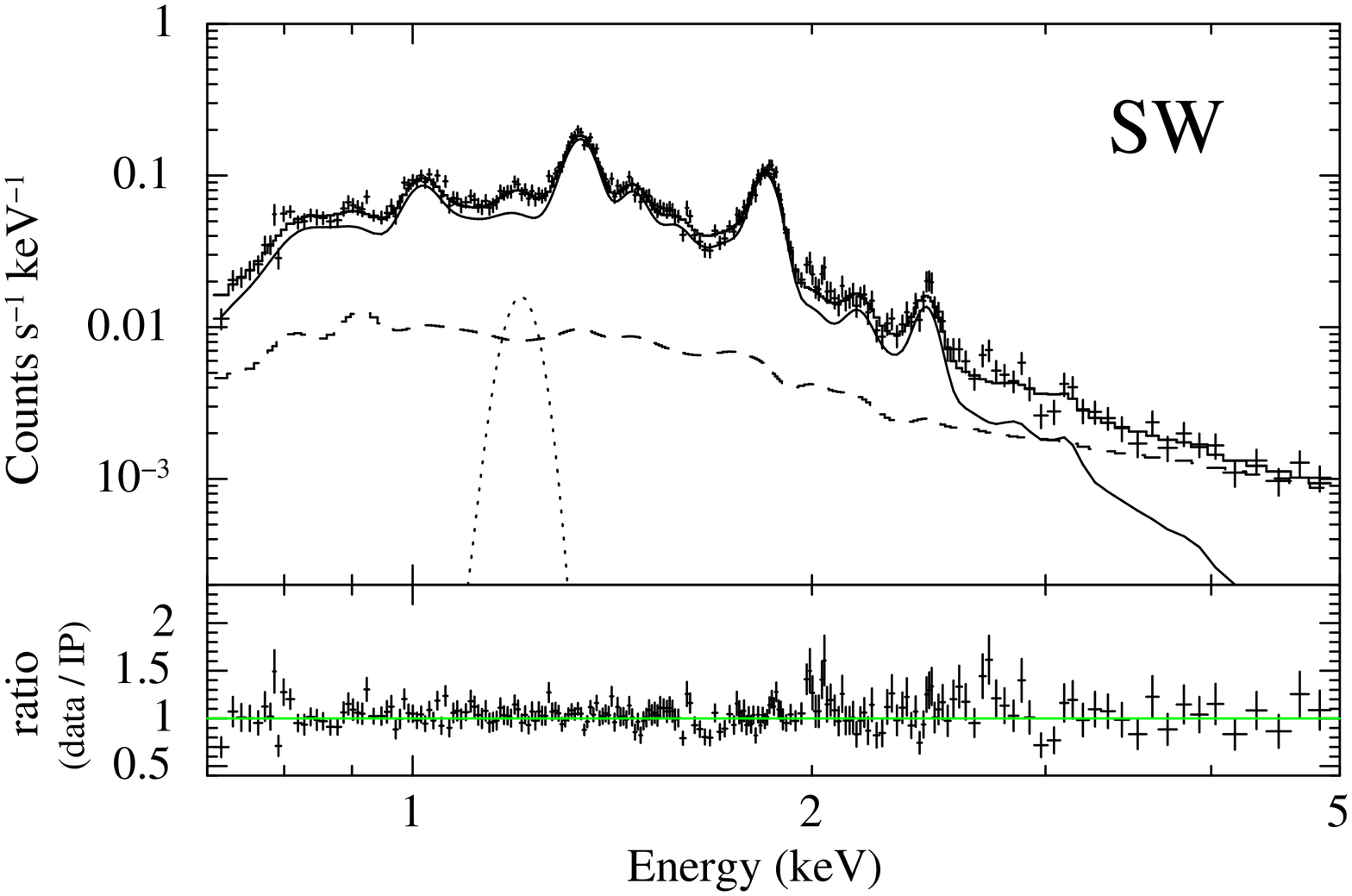}
	\end{center}
	\end{minipage}
	\begin{minipage}{0.5\hsize}
	\begin{center}
		\FigureFile(80mm,80mm){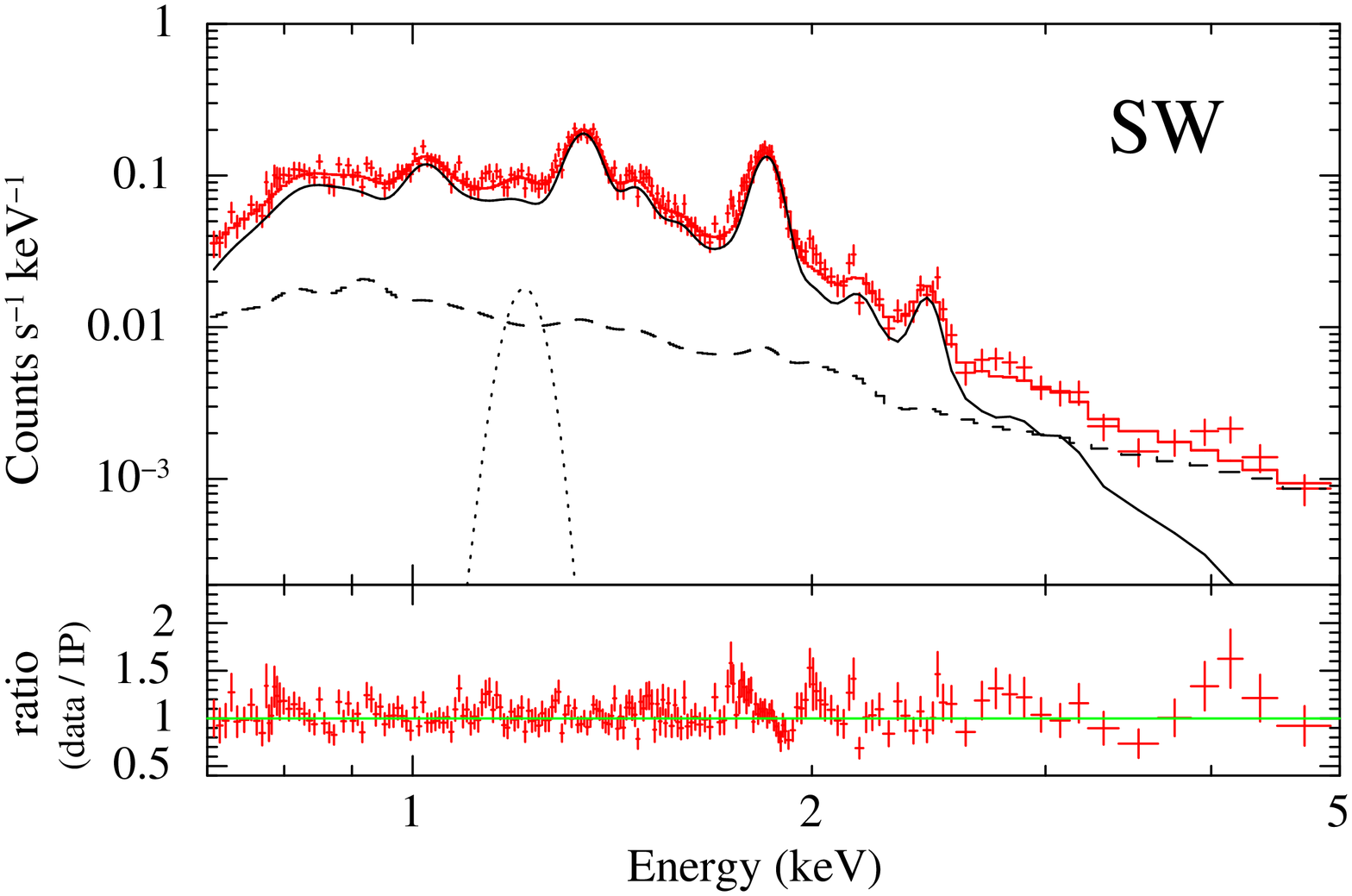}
	\end{center}
	\end{minipage}
	\caption{Continued.}
\label{spectra}
\end{figure}

\newpage
\begin{table*}
\caption{Best-fit parameters for G290.1$-$0.8 spectra.}\label{fit-para}
\begin{center}
\begin{tabular}{lcccccc}
\hline\hline
Parameter & & Center & NW & SE & NE & SW \\
\hline
 & & RP & RP & IP & IP & IP\\
\hline
$N_{\rm H}$ $(10^{21} {\rm cm^{-2}})$ & & $7.8_{-0.5}^{+0.3}$ & $9.6_{-1.4}^{+0.8}$ &  $6.8_{-0.8}^{+0.7}$ & $6.7_{-0.6}^{+0.5}$ &  $6.2_{-0.8}^{+0.7}$\\
$kT_{\rm init}$ (keV) & & $1.7_{-0.3}^{+1.2}$ & $>1.9$ & $-$ & $-$ & $-$ \\
$kT_{\rm e}$ (keV) & & $0.45_{-0.01}^{+0.02}$ & $0.52_{-0.04}^{+0.02}$ & $0.66_{-0.02}^{+0.03}$ & $0.64_{-0.01}^{+0.02}$ & $0.59_{-0.02}^{+0.03}$\\
$n_{\rm e} t$ $(10^{12} {\rm s\,cm^{-3}})$ & & $1.22_{-0.09}^{+0.13}$ & $1.06_{-0.20}^{+0.06}$ & $0.34_{-0.06}^{+0.07}$ & $0.42_{-0.07}^{+0.09}$ & $0.64_{-0.18}^{+0.25}$\\
$EM^\dag$ $(10^{11} {\rm cm^{-5}})$ & & $23_{-1}^{+2}$ & $4.0_{-0.5}^{+0.7}$ &  $4.9_{-0.5}^{+0.4}$ & $5.9\pm0.4$ &  $4.9_{-0.5}^{+0.3}$\\
Ne$^\ddag$ & & $0.43_{-0.06}^{+0.04}$ & $0.7\pm0.2$ & $0.36_{-0.07}^{+0.08}$ & $0.43_{-0.08}^{+0.09}$ & $0.5\pm0.1$\\
Mg$^\ddag$ & & $1.5\pm0.1$ & $1.5\pm0.3$ & $1.3\pm0.1$ & $1.2\pm0.1$ & $1.2\pm0.2$\\
Si$^\ddag$ & & $2.5_{-0.2}^{+0.1}$ & $2.6_{-0.3}^{+0.4}$ & $2.0_{-0.1}^{+0.2}$ & $1.6\pm0.1$ & $1.4_{-0.1}^{+0.2}$\\
S$^\ddag$ & & $2.1\pm0.2$ & $2.4_{-0.3}^{+0.4}$ & $1.5_{-0.2}^{+0.1}$ & $1.1\pm0.1$ & $1.1_{-0.2}^{+0.1}$\\
Ar$^\ddag$ (= Ca$^\ddag$) & & $2\pm1$ & $3\pm1$ & $1.1_{-0.6}^{+0.7}$ & $1.3\pm0.6$  & $< 1.6$\\
Fe$^\ddag$ (=Ni$^\ddag$) & & $0.10_{-0.03}^{+0.02}$ & $0.3_{-0.2}^{+0.1}$ & $0.16_{-0.05}^{+0.06}$ & $0.22\pm0.05$ & $0.18_{-0.06}^{+0.07}$\\
\hline
$\chi^2_{\nu} / {\rm d.o.f.}$ & & $1.12/1436$ & $1.10/749$ &  $1.08/1126$ & $1.06/1198$ & $1.10/1080$\\
\hline
\multicolumn{7}{l}{\small $^\dag$ Volume emission measure, $\int n_e n_{\rm H} dV / (4 \pi D^2)$, where {\it{V}} and {\it{D}} are the emitting volume (cm$^3$) and}\\[-1.5mm]
\multicolumn{7}{l}{\small the distance to the source (cm), respectively.}\\[-1.5mm]
\multicolumn{7}{l}{\small $^\ddag$ Relative to the solar value (\cite{Anders1989}).}\\
\end{tabular}
\end{center}
\end{table*}

\begin{figure}
   \begin{center}
         \FigureFile(80mm,80mm){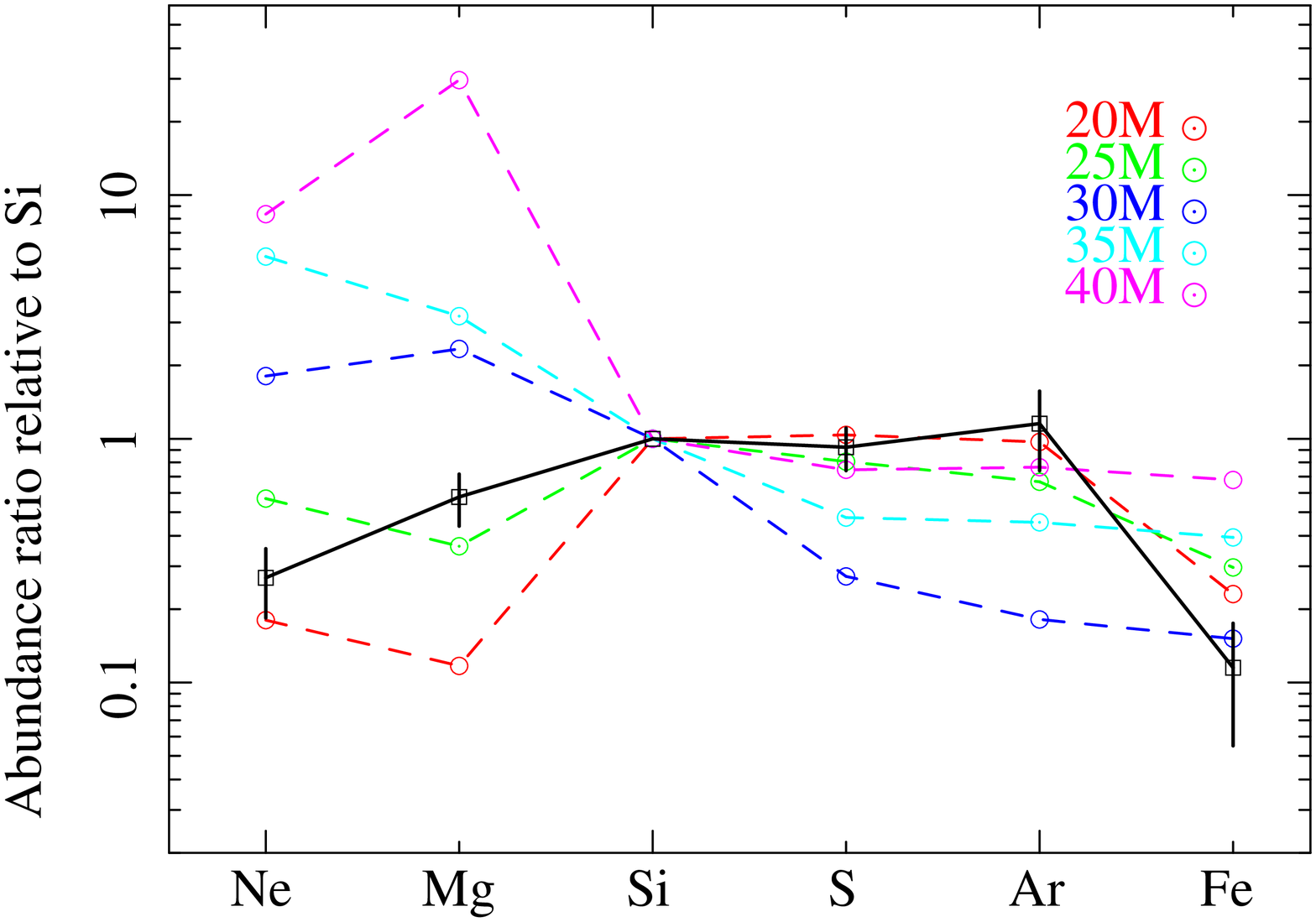}
   \end{center}
   \caption{Abundance pattern of the NW plasma (black solid line). Each elemental abundance is normalized by the value of Si. The red, green, blue, light blue, and magenta dashed lines represent core-collapse models with progenitor masses of 20, 25, 30, 35, and 40 $M_{\odot}$, respectively (\cite{Woosley1995})}
\label{abund_pattern} 
\end{figure}

\end{document}